# Glass shaping at nanoscale: Mechanical forming of brittle amorphous silica by engineered inelastic interaction of scanning electrons with matter

Sung-gyu Kang[1], Kyeongjae Jeong[1], Woo Jin Cho[1], Jeongin Paeng[1], Jae-Pyeong Ahn[2], Steven Boles[3] and Heung Nam Han[1*], In-Suk Choi[1*]

[1]Department of Materials Science and Engineering & Research Institute of Advanced Materials, Seoul National University, Seoul, Republic of Korea

[2]Advanced Analysis Center, Korea Institute of Science and Technology, Seoul, Republic of Korea

[3]Department of Electrical Engineering, The Hong Kong Polytechnic University, Hung Hom, Kowloon, Hong Kong

E-mail addresses: hnhan@snu.ac.kr (H.N. Han), insukchoi@snu.ac.kr (I.-S. Choi)

**ABSTRACT**

Amorphous silica deforms viscoplastically at elevated temperatures, as is common for brittle glasses. The key mechanism of viscoplastic deformation involves interatomic bond switching, which is known to be a thermally activated process. In this study, through systematic *in-situ* compression tests by scanning electron microscopy, the viscoplastic deformation of amorphous silica is observed without thermal activation. Furthermore, ductility does not increase monotonically with acceleration voltage and current density of the SEM e-beam but is maximized by a factor of three at a specific acceleration voltage and current density conditions (compared to beam-off conditions). A Monte Carlo simulation of the electron-matter interaction shows that the unique trends of viscoplastic deformation correlate with the interaction volume, i.e., the region within the material where inelastic electron scattering occurs. Changing the size of the migrating atomic clusters can lead to facility in rearrangements of the intramolecular bonds, hence leading to more sustained bond switching. Based on the interaction volume the mechanical shaping of small-scale amorphous silica structures under e-beam irradiation can be modeled with high-precision supporting the idea that this relatively low-voltage e-beam-irradiation induced viscoplastic-deformation technique holds great potential for advancing amorphous silica structure manufacturing and developing e-beam assisted manufacturing for covalently bonded non-metallic materials.

Glass, a very hard and chemically stable material, has seen increasingly ubiquitous use since ancient times because of its formability. Under high temperatures, glass is easily transformed into useful shapes using a wide range of glass crafting techniques. Its mechanically formable nature is due to viscoplastic deformation of the amorphous material under high temperatures (exceeding the glass transition temperature, $T_g$). In this study, we demonstrate the mechanical shaping of amorphous silica on a small scale (from nano- to micrometer) at room temperature based on the inelastic scattering interaction between electrons and matter by utilizing a focused electron beam (henceforth referred to as e-beam) with low acceleration voltage. The ductile super-plastic deformation behavior of amorphous silica under e-beam irradiation is directly related to the electron-matter interaction volume. Experiments and simulations have shown that transmitted and scattered electrons can alter the bond nature of amorphous silica.

Amorphous silica mainly consists of strong covalently bonded silicon and oxygen atoms which are abundant elements in the earth's crust. The strong covalent bonds are switchable at temperatures exceeding $T_g$ and only with such conditions does amorphous silica deform viscoplastically, exhibiting formability. Although small-scale amorphous silica has become integral in advanced electronics [1, 2, 3, 4, 5, 6, 7], biomaterials [8, 9, 10, 11, 12], and 3D nano-structural materials [13, 14, 15], its temperature-dependent formability limits process compatibility. Researchers have attempted to improve formability of material by applying electric current. During the Electrically-Assisted-Manufacturing (EAM), the electroplasticity improves the formability of materials but occurs only in metals [16, 17, 18, 19]. Several studies have tried to apply electric current to the sintering process of glass material, but there has been no significant improvement in formability [20, 21]. Meanwhile, recent studies have shown that e-beam irradiation can induce ductile super-plastic deformation of amorphous silica [22, 23, 24, 25, 26, 27, 28], which may have enormous potential for utilization in small-scale fabrication processes by using current e-beam technologies [29, 30, 31, 32]. However, applying this unique e-beam induced ductile super-plastic behavior remains still a significant challenge because the community has a very limited understanding of how electron-matter interaction affect the deformation behavior of amorphous silica under e-beam irradiation. For an in-depth understanding and practical application, we believe that the irradiation effects of a wide range of e-beam conditions (especially with low acceleration voltages from a few to tens of kV without causing any e-beam damage) on amorphous silica must be studied. To this end, we systematically perform *in-situ* scanning electron microscopy (SEM) compression tests by modulating parameters such as the acceleration voltage, current density, and spatial area of irradiation (whereas other studies have used fixed high acceleration voltage (hundreds of kV) e-beams of conventional transmission electron microscopies (CTEM) [22, 23, 24, 25, 26, 27, 28]). Consequently, we are able to systematically analyze e-beam induced super-plastic behavior in terms of the interaction between the incident electrons and matter with help of Monte-Carlo simulations. Finally, based on this understanding, we demonstrate the feasibility of mechanical shaping of small-scale amorphous silica under scanning e-beam irradiation.

**Unique super-plastic deformation behavior of amorphous silica nano-sphere under low-voltage scanning e-beam irradiation**

We started by synthesizing amorphous silica spheres through a sol-gel method, the Stöber process (Supporting Information S1). The synthesized amorphous silica spheres exhibit a uniform spherical geometry with a narrow diameter distribution (average diameter of 290 nm, Figure S1). Moreover, they are free of irradiation damage, unlike silica spheres synthesized through an ion-beam milling process. We investigated the effect of e-beam irradiation on the deformation behavior of the amorphous silica spheres through *in-situ* SEM compression tests using a 500-nm-radius cono-spherical indenter (Supporting Information S2 and Figure S2).

Scanning e-beams with low acceleration voltages clearly induce the immediate super-plastic deformation of amorphous silica spheres under compression despite completely different characteristic of e-beam irradiation from the previous CTEM studies. Figure 1 shows how the deformation behavior changes under e-beam irradiation with an acceleration voltage of 5 kV and current density of 27.9 A/m$^2$. In the absence of e-beam irradiation (beam-off conditions), most deformation is elastic and minimal plastic deformation occurs. Figure 1 (a-1) shows the load-depth curve (black) of the compression test under an applied load of 100 μN and beam-off conditions. The maximum compression depth is 43 nm, but the residual compression depth is only 10 nm due to the subsequent elastic recovery. There is little difference in the shape of the amorphous silica sphere before (Figure 1 (a-2)) and after compression (Figure 1 (a-3)). It is estimated that limited plastic deformation under compression is due to a densification of the amorphous silica [33, 34, 35]. Then, the compression tests under larger applied loads (e.g., 300 μN) induce brittle fracturing of the amorphous silica, which is signaled by a meridian crack on the surface of the sphere (Figure 1 (a-4)). Pop-in events identified in the compression load-depth curve in Figure S3 are consistent with corresponding crack formation. However, under scanning e-beam irradiation, the amorphous silica sphere becomes ductile and accommodates significant permanent plastic deformation. Figure 1 (a-1) shows the compression load-depth curve (green) under an applied load of 100 μN and e-beam irradiation. The curve is less steep than that under beam-off conditions and the maximum compression depth reaches 149 nm. A series of snapshots (Figure 1 (a-5)) and Movie S1 show the continuous deformation of an amorphous silica sphere beneath the indenter until the maximum compression depth is achieved. The residual compression depth after unloading is 140 nm, indicating that the deformation is mostly plastic. This is confirmed by the pancake-like compressed shape of the amorphous silica sphere in Figure 1 (a-6). Interestingly, no cracking on the sphere surface, even after the significant plastic deformation, is observed. Moreover, a mixed-mode compression test (i.e. performed while alternating between beam-off and beam-on conditions) suggests that the amorphous silica sphere immediately transitions between ductile and stiff responses depending on the presence of e-beam irradiation. As per the compression load-depth curve in Figure 1 (b) and corresponding *in-situ* Movie S2, the slope decreases by a factor of 4 when the e-beam is turned on. Conversely, when we turn the e-beam off, the slope increases back close to original shape. In other words, the brittle amorphous silica sphere becomes ductile immediately upon e-beam irradiation. Similar e-beam irradiation induced ductile super-plastic deformation of amorphous silica has been reported in a study using a CTEM e-beam with an acceleration voltage of 200–300 keV to irradiate an entire sphere [28]. It should be noted that our research conducted using SEM demonstrates that the same ductile super-plastic deformation appears even when a spatially focused e-beam with an acceleration voltage 40 times lower than that of the TEM e-beam scans the amorphous silica sphere following the raster pattern.

Subsequent tests found that the super-plastic deformation of the amorphous silica spheres shows unique dependencies on the acceleration voltage ($V_A$) and the current density ($J$) of the e-beam. We systematically conducted compression tests on the amorphous silica spheres under applied loads of 100 μN and e-beam irradiation while varying $V_A$ and $J$ from 1 to 30 kV and 0.43 A/m² to 27.94 A/m², respectively (Table S1). The SEM images of the amorphous silica spheres after compression in Figure 2 (a) simply show that the higher the $V_A$ or $J$ of the e-beam, the greater the super-plastic deformation of the amorphous silica sphere during compression. The compressed shapes of the amorphous silica spheres at $V_A$ = 1 kV or $J$ = 0.43 A/m² (first column or first row of SEM images in Figure 2 (a)) are similar to the shape obtained under beam-off conditions (Figure 1 (a-3)). The SEM images enclosed in the red box in Figure 2 (a) show that as $V_A$ or $J$ increases the compressed amorphous silica sphere gradually changes into a pancake-like shape. However, a quantitative analysis, focusing on the compression load-depth curves in Figure S4, reveals the unique $V_A$ and $J$ dependency of the super-plastic deformation of the amorphous silica spheres. Considering the self-similarity of the amorphous silica spheres, we calculated the nominal flow stress (dividing the compression load by $\pi r^2$) at 0.1 nominal strain (i.e., where the compression depth is 10 % of the sphere diameter) from the compression load-depth curves. By dividing the nominal flow stress under each e-beam condition ($\sigma$) by that under beam-off conditions ($\sigma_{beam-off}$), we define the normalized flow stress ($\xi_\sigma$) as below,

$$\xi_\sigma = \sigma/\sigma_{beam-off} \tag{1}$$

which represents the flow stress change under the e-beam. For example, under beam-off conditions, the $\xi_\sigma$ of the amorphous silica sphere should equal 1 and as $\xi_\sigma$ decreases, super-plastic deformation during compression increases.

Our quantitative analysis, as per the $\xi_\sigma$ versus $J$ plot for different $V_A$ (Figure 2 (b)), produced three major findings. First, a $V_A$ threshold exists, above which e-beam irradiation induces super-plastic deformation. As mentioned, under beam-off conditions, $\xi_\sigma = 1 \pm 0.12$ (black square in Figure 2 (b)). While, at other $V_A$ under e-beam irradiation, $\xi_\sigma < 1$ (orange, gold, green, light-blue, blue, and purple squares in Figure 2 (b)), $\xi_\sigma \approx 1$ only at $V_A = 1$ kV (red squares in Figure 2 (b)). Thus, $V_A$ should exceed 1 kV to induce super-plastic deformation in the amorphous silica spheres. Second, $\xi_\sigma$ does not decrease monotonically with $J$, and becomes plateau when $J$ exceeds 6.98 A/m². For example, at $V_A = 5$ kV, $\xi_\sigma$ initially decreases as $J$ increases and then becomes constant at $\xi_\sigma = 0.33$ (green squares in Figure 2 (b)). Thus, the amorphous silica spheres experience similar super-plastic deformation when $J > 6.98$ A/m². This trend is repeated for other values of $V_A$ (i.e., 2, 3, 10, 20, and 30 kV) in Figure 2 (b). Finally, the amorphous silica spheres go through the largest super-plastic deformation under a 5-kV e-beam. At every value of $J$ (except for 0.43 A/m$^2$), $\xi_\sigma$ decreases when $V_A > 1$ kV but records the lowest value at $V_A = 5$ kV (green square in Figure 2 (b)). At $V_A = 5$ kV and $J = 27.94$ A/m$^2$, $\xi_\sigma = 0.32$, indicating that the flow stress is 68 % less than that under beam-off conditions. When $V_A$ exceeds 5 kV, $\xi_\sigma > 0.32$ and $\xi_\sigma$ at $V_A = 30$ kV (purple squares in Figure 2(b)) is even comparable to that of $V_A = 3$ kV (gold squares in Figure 2 (b)). A previous research work about e-beam irradiation damage of materials indicates that the damage increases with increasing particle energy, which correlates with the power of the e-beam [36]. In this context, we initially considered the possibility that the ductility should monotonically increase with $V_A$ and $J$ of the e-beam. Therefore, our observation that optimal values of $V_A$ and $J$ exist, that maximizes the ductility of amorphous silica spheres, may be counterintuitive. However, simulation of the electron-matter interaction offers insights into these findings regarding e-beam irradiation induced deformation of amorphous silica spheres.

**Role of electron-matter interaction in superplastic behavior under e-beam irradiation**

When a solid material is irradiated with accelerated electrons, there is an interaction between electrons and matter, which often leads to changes in the material properties [31, 37]. The electron-matter interaction can be described using a Monte Carlo simulation (CASINO™ software (Ver. 3.3)) based on elastic and inelastic scattering events [38, 39]. Considering the lifetimes of the excited electrons, the settling times of the charges produced by the electron-matter interaction, and the scanning profile of the e-beam, we constructed a model simulating the interaction during a single line scan by the e-beam across the amorphous silica sphere [40] (Supporting Information S3 and Figure S5). Figure 3 (a) shows a side view and a top view of the interaction model. Green arrows and line indicate a single line scan profile of the e-beam. From elastic collisions, the incident electrons can be transmitted or reflected within the material.

Alternatively, through the inelastic collisions, the incident electrons transfer energy to the solid matter. The transferred energy possibly changes the nature of the interatomic covalent bond, making it switchable. As Monte Carlo model simulates the energies and trajectories of the incident electrons, we can obtain the total energy that the system absorbs during the inelastic scattering event and the total volume of the region where the inelastic scattering event occurs. Accordingly, we investigated whether these two values explain the above experimental findings about the super-plastic deformation in amorphous silica spheres depending on $V_A$ and $J$ of the scanning e-beam.

At first, the total absorbed energy ($E_{total}$) does not explain the maximum super-plastic deformation of amorphous silica observed under an e-beam with $V_A = 5$ kV and $J = 27.94$ A/m². Figure 3 (b) shows the total absorbed energy with respect to $J$ for different $V_A$. The changes observed in $E_{total}$ is very different than that of $\xi_\sigma$ in Figure 2 (b). $E_{total}$ is simply proportional to $J$ at every $V_A$ while $\xi_\sigma$ is inversely proportional and then becomes constant at J > 6.98 A/m². In addition, at every $J$, $E_{total}$ records the highest and the lowest values at $V_A = 30$ and 3 kV, respectively, and values differ by one order of magnitude (purple-gray squares and gold-gray squares in Figure 3 (b)). On the other hand, as shown in Figure 2 (b), at $V_A = 30$ and 3 kV, $\xi_\sigma$ is almost same for all $J$ (purple squares and gold squares in Figure 2 (b)). Moreover, $E_{total}$ at $V_A = $ 10 kV is comparable to that at $V_A = 1$ kV, which is inconsistent with the $V_A$ threshold determined for super-plastic deformation (light blue-gray and red-gray squares in Figure 3 (b), respectively). The distribution of absorbed energy ($E_a$) also deviates from the deformation trends observed in *in-situ* experiments on the amorphous silica spheres. The energy distributions along and normal to the single line scan profile of e-beam in Figure 3 (c) illustrate that most of the energy is absorbed at the surface. In the case of $V_A = 3$ kV and $J = 27.9$ A/m² (intersection of second row and third and sixth columns in Figure 3 (c)), $E_a$ at the irradiated surface is almost 10 times greater than that at 50 nm beneath the surface. If the super-plastic deformation is directly related to the absorbed energy, most deformation should occur around the irradiated surface and the deformed shape should be asymmetric. However, in the actual experiment under the e-beam with the same conditions, the amorphous silica sphere is compressed symmetrically (intersection of fourth row and third column in Figure 2 (a)). The apparent disconnect between $E_a$ and $\xi_\sigma$ signifies that the super-plastic deformation of the amorphous silica sphere does not simply depend on the energy absorbed from the e-beam.

In contrast, the interaction volume ($\Omega_I$), i.e., the entire volume of the region where the energy absorption occurs through an inelastic scattering event, shows very strong correlation with $\xi_\sigma$. This is illustrated by the non-interaction volume fraction ($f_{\Omega_{NI}}$) which is defined as,

$$f_{\Omega_{NI}} = \frac{\Omega_{NI}}{\Omega_{sphere}} = \frac{\Omega_{sphere} - \Omega_I}{\Omega_{sphere}} \tag{2}$$

where $\Omega_{sphere}$ and $\Omega_{NI}$ indicate the volume of the amorphous silica sphere and the non-interaction volume inside the sphere (where energy absorption does not occur), respectively. For example, under beam-off conditions, the $f_{\Omega_{NI}}$ of the amorphous silica sphere should be 1 and $f_{\Omega_{NI}}$ decreases as $\Omega_I$ increases. From the $f_{\Omega_{NI}}$ versus $J$ plot for different $V_A$ in Figure 3 (d), three characteristics can be identified, which are very similar to those findings of $\xi_\sigma$ versus $J$ plot (Figure 2 (b)). First, $f_{\Omega_{NI}} = 0.95 \pm 0.03$ at $V_A = 1$ kV regardless of $J$, which is no different than that under beam-off conditions (red squares in Figure 3 (d)). Second, except for $V_A = 1$ kV, $f_{\Omega_{NI}}$ decreases at every $V_A$ with increasing $J$ and then becomes constant when $J$ exceeds 6.9 A/m$^2$. Third, between the various $V_A$, $f_{\Omega_{NI}}$ records the lowest value at $V_A = 5$ kV. The characteristic dependency of $f_{\Omega_{NI}}$ on $V_A$ and $J$ are surprisingly similar to that of $\xi_\sigma$, i.e. the super-plastic deformation of the amorphous silica sphere. Simulated cross-section images of $\Omega_I$ along and normal to the single line scan profile of e-beam in Figure 3 (e) provides a physical explanation of the correlation between $f_{\Omega_{NI}}$ and $\xi_\sigma$. First, at $V_A = 1$ kV, $f_{\Omega_{NI}} \approx 1$ because the incident electrons penetrate only 40 nm (losing all their energy) and the $\Omega_I$ cross sections show that only a limited volume of matter near the irradiated surface interacts with the incident electrons (denoted by an orange color). Since there is no change in the deformation behavior of the region where the interaction does not occur, $\xi_\sigma$ at $V_A = 1$ kV is almost the same as that under beam-off conditions. Second, a constant $f_{\Omega NI}$ at $J > 6.9$ A/m² is due to a saturation of $\Omega_I$ inside the amorphous silica sphere. This is explained by comparing the $\Omega_I$ cross sections along the single line scan profile of e-beam at $J = 1.8, 6.9,$ and $27.9$ A/m² in Figure 3 (e). At $J \leq 6.9$ A/m², $\Omega_I$ expands with increasing $J$ for every $V_A$. Notably, at $J = 6.9$ A/m², $\Omega_I$ already covers most of the volume of the sphere from the surface to the depth to which the incident electrons can penetrate. A further increase in $J$ does not leads to $\Omega_I$ expansion but its saturation, which means that $\xi_\sigma$ will not decrease further and will become constant. Third, between the various $V_A$, $f_{\Omega NI}$ records the lowest value at $V_A = 5$ kV because the incident electrons are scattered and radially spread inside the amorphous silica sphere, resulting in the interaction occurring in most of $\Omega_{sphere}$. It is attributed to the increasing penetration depth and decreasing scattering angle of electron as $V_A$ increases. At $V_A > 1$ kV, $f_{\Omega NI}$ decreases because the incident electrons penetrate deeper and the $\Omega_I$ cross-section images at $V_A = 3$ kV and $J = 27.9$ A/m² (intersections of the second row with the third and sixth columns in Figure 3 (e)) show that the interaction occurs up to a 140 nm depth from the irradiated surface. Especially, at $V_A = 5$ kV, the incident electrons have sufficient energy to pass through the amorphous silica sphere. Therefore, $\Omega_I$ at $V_A = 5$ kV and $J = 27.9$ A/m² covers the cross-section images along and normal to the single line scan profile of e-beam (intersections of the third row with third and sixth columns in Figure 3 (e)) and corresponding $f_{\Omega NI}$ is minimized and becomes as small as 0.04. The increase in $\Omega_I$ with $V_A$ makes the amorphous silica sphere more ductile and eventually $\xi_\sigma$ decreases. However, further increase in $V_A$ decreases $\Omega_I$ so that $f_{\Omega NI}$ increases at $V_A > 5$ kV, as the incident electrons can pass through the sphere with a narrower scattering angle. It is evident from the $\Omega_I$ cross-section images of $V_A = 5, 10,$ and $30$ kV in Figure 3 (e). At $J = 27.9$ A/m², $\Omega_I$ of $V_A = 5, 10,$ and $30$ kV commonly cover the whole cross-section image along the single line scan profile of e-beam, indicating the penetration depth is greater than the sphere diameter (intersections of third to fifth rows with third column in Figure 3 (e)). However, in the cross-section image normal to the single line scan profile of e-beam, $Ω_I$ gradually decreases as $V_A$ increases from 5 to 30 kV (intersections of third to fifth rows with sixth column in Figure 3 (e)). This is because the incident electrons accelerated by 5 kV deviate significantly from their original trajectory during the scattering events and radially spread inside the amorphous silica sphere, whereas the incident electrons accelerated by 10 and 30 kV deviate from their original trajectory at a relatively small angle during the scattering events and directly penetrate towards the opposite side of the sphere. Subsequently, $Ω_{NI}$ (denoted in white) increases in the cross-section images normal to the single line scan profile of e-beam at $V_A > 5$ kV, which corresponds with an increase in $ξ_σ$. The strong correlation between $Ω_I$ and $ξ_σ$ suggests that the counter-intuitive $V_A$ and J dependency of the super-plastic deformation of the amorphous silica sphere can be ascribed to the trajectory of the incident electrons and $Ω_I$ generated therefrom. This also explain how a focused scanning e-beam can induced super-plastic deformation even though the beam spot is much smaller than that of the previous CTEM e-beam conditions.

**Strain-rate dependency of e-beam induced super-plasticity**

The deformation behavior of an amorphous silica pillar, in which a uniform stress field develops during compression, also heavily depends on $Ω_I$. We fabricated pillars with a diameter of 280 nm and a height of 900 nm on a fused quartz substrate through an ion-beam milling process (Supporting Information S4) and conducted *in-situ* compression tests on them with a 500-nm-radius flat punch indenter under beam-off; 2-kV, 1.6-A/m$^2$; 5-kV, 1.4-A/m$^2$; 18-kV, 2.0-A/m$^2$; and 30-kV, 2.4-A/m$^2$ e-beam conditions (Supporting Information S5 and Figure S6).

Figure 4 (a) shows the $\xi_\sigma$ from the *in-situ* compression tests of the fused quartz pillars and corresponding $f_{\Omega NI}$ determined using the Monte Carlo simulation. The model simulates the interaction inside the fused quartz pillar provoked by the single line scan of SEM e-beams with $V_A$ = 2, 5, 18, and 30 kV and $J$ = 1.8 A/m² (Supporting Information S6 and Figure S7). The $\xi_\sigma$ and $f_{\Omega NI}$ of the fused quartz pillar also exhibit similar trends, consistent with the trends observed for the amorphous silica spheres. Both $\xi_\sigma$ and $f_{\Omega NI}$ of the fused quartz pillar achieve a minimum value at $V_A$ = 5 kV. The $\Omega_I$ cross-section images normal to the single line scan profile of e-beam in Figure 4 (b) show the increasing penetration depth and the decreasing scattering angle of the incident electrons with increasing $V_A$. The $\Omega_I$ is maximized at $V_A$ = 5 kV. Even for the fused quartz pillar, $\Omega_I$ explains the $V_A$ dependence of the super-plastic deformation behavior well. Furthermore, it appears that the super-plastic deformation of amorphous silica occurs under e-beam irradiation regardless of the fabrication method of silica nanostructures.

To address the question of how amorphous silica deforms when inelastic scattering of incident electrons occurs within the material, we investigated the strain rate sensitivity, a parameter of the deformation mechanism of materials [41, 42, 43]. The strain rate sensitivity (*m*) is defined as,

$$m = \frac{\partial \ln \sigma}{\partial \ln \dot{\varepsilon}} \tag{3}$$

where $\sigma$ is the nominal flow stress at 0.1 nominal strain, and $\dot{\varepsilon}$ is the strain rate. For amorphous materials, a value of *m* = 0 indicates a rigid plastic material, while *m* = 1 indicates a linear viscous solid [42, 43]. To calculate the *m* of amorphous silica under e-beam irradiation, we conducted *in-situ* compression tests on fused quartz pillars by varying $\dot{\varepsilon}$ from 6.7×10⁻⁴ s⁻¹ to 1.4×10⁻² s⁻¹ under beam-off; 2-kV, 1.6-A/m²; 5-kV, 1.4-A/m²; and 18-kV, 2.0-A/m² e-beam conditions (Supporting Information S7 and Figure S8).

The fused quartz pillars interacting with the incident electrons undergo viscoplastic deformation. This is evident from the slopes (i.e. *m*) of the $\sigma$-$\dot{\varepsilon}$ curves of the fused quartz pillars under various e-beam conditions (Figure 4 (c)). Under beam-off conditions, the deformation of the fused quartz pillar is mostly rigid plastic. The black squares and dotted line in Figure 4 (c) show that $\sigma$ is almost independent of $\dot{\varepsilon}$ and the corresponding *m* is as small as $5.98\times10^{-3}$, which is consistent with other results measured under ambient conditions [43, 44, 45]. Under e-beam irradiation, the deformation becomes viscoplastic. Under 2-kV, 1.6-A/m$^2$; 5-kV, 1.4-A/m$^2$; and 18-kV, 2.0-A/m$^2$ e-beams, $\sigma$ increases with $\dot{\varepsilon}$ (respectively denoted by orange, green, and blue squares in Figure 4 (c)). The *m* under 2-kV, 1.6-A/m$^2$ and 18-kV, 2.0-A/m$^2$ e-beams increase to $7.93\times10^{-2}$ and $1.77\times10^{-1}$, respectively (orange and blue dotted lines in Figure 4 (c)). Under a 5-kV, 1.4-A/m$^2$ e-beam, *m* records $2.71\times10^{-1}$, a dramatic 45-fold increase from that under beam-off conditions (green dotted line in Figure 4 (c)). Notable increases in *m* of the fused quartz pillar under e-beam irradiation indicates that the amorphous silica deforms viscoplastically during compression.

The viscoplastic deformation of amorphous material is known to occur at high temperatures exceeding $T_g$ ($T_g$ of amorphous silica is 1200 °C), owing to the fact that interatomic bond switching, a key mechanism of the deformation, is a thermally activated process [22, 46]. However, the viscoplastic deformation of amorphous silica in our study is not thermally activated. We calculated $E_{total}$ of the fused quartz pillar under a 5-kV, 1.4-A/m$^2$ e-beam and then estimated a temperature rise (Supporting Information S8 and Figure S9). From a finite element (FE) simulation, which assumes that $E_{total}$ is concentrated in a limited volume and is fully converted into a thermal energy, the temperature rise induced under a 5-kV, 1.4-A/m$^2$ e-beam is less than 1 °C (Figure S10). Considering that the simulation overestimates the temperature rise, the viscoplastic deformation of amorphous silica under e-beam irradiation can be considered as an athermal process. This implies that the incident electrons in our study directly affect the interatomic bond switching process and bring about viscoplastic deformation of amorphous silica. The atoms can overcome the energy barrier to interatomic bond switching more easily when the interatomic bond nature changes. Depending on the acceleration voltage, incident electrons affect materials through elastic and inelastic scattering. Through elastic scattering, a charged particle can produce a vacancy in a material. This vacancy formation, or so-called knock-on damage, is not possible in our low acceleration voltage irradiation because it requires the incident electron accelerated by hundreds of MV [22, 23, 24, 40, 47]. Meanwhile, through inelastic scattering, incident electrons accelerated by few to tens of kV induces ionization and secondary electron emission [40, 47], which possibly changes the interatomic bond nature and makes the interatomic bond more switchable. Monte Carlo simulation indicates that the incident electrons with low $V_A$, from 1 to 30 kV, yield secondary electrons in the fused quartz pillar (Supporting Information S9 and Figure S11). It suggests that the ionization occurs in the fused quartz pillar and, concomitantly, the interatomic bond nature changes similar to that of high temperature condition. The activation volume ($V^*$), i.e., the volume of an atom cluster migrating during the deformation, supports this conclusion. From the strain rate sensitivity analysis, we calculate $V^*$ as below,

$$V^* = \sqrt{3}kT \left(\frac{\partial \ln \dot{\varepsilon}}{\partial \sigma_n}\right) \tag{4}$$

where $k$ is Boltzmann's constant and $T$ is the temperature. The lower the $V^*$ of the amorphous material, the more viscoplastic is its deformation. We assumed $T$ to be 298 K because there is almost no temperature rise induced by e-beam irradiation. Figure 4 (d) shows the $\dot{\varepsilon}$ - $\sigma$ curves of the fused quartz pillars under various e-beams with corresponding $V^*$. In the absence of e-beam irradiation, the activation volume is approximately 138.26 Å$^3$ which is consistent with the value at ambient temperature [45]. Under e-beam irradiation, the activation volume drastically decreases. Especially, under a 5-kV, 1.4-A/m$^2$ e-beam, the activation volume is 9.64 Å$^3$ corresponding to an atom cluster consisting of 2–3 atoms (or approximately one molecular formula unit) that is consistent with the value at high temperature [48, 49, 50]. It implies that e-beam irradiation makes the interatomic bond easier to switch in an athermal way based on the fact that the activation volume decreases just as in the thermally activated deformation of glass at high temperature. In summary, every single incident electron has sufficient energy to induce the ionization and change the interatomic bond nature to make these bonds more switchable. Accordingly, the amorphous silica becomes ductile and undergoes viscoplastic deformation similar to thermally activated viscoplastic deformation. For the same reason, the viscoplastic deformation of the amorphous silica strongly depends on the volume of the region where the ionization occurs, that is, $\Omega_I$, not $E_a$.

**Engineering guidance of mechanical shaping based on interaction volume**

This discovery of e-beam irradiation induced deformation of brittle amorphous silica may offer a new approach to mechanical shaping. We demonstrate that we can successfully model the various shape forming of amorphous silica under scanning e-beam irradiation using FE analysis. First, we determined the mechanical properties of the amorphous silica for constitutive equations, during full interaction with incident electrons (5-kV, 27.9-A/m$^2$) and in the absence of incident electrons (beam off), via reverse engineering which minimizes the difference between load-depth curves obtained from experimental and computational compression tests of amorphous silica spheres (Supporting Information S10 and Figure S12). Then, we treated the amorphous silica subjected to scanning e-beam irradiation as a composite of $\Omega_{NI}$ and $\Omega_I$. For instance, we could model the localized shape forming of amorphous silica in which a volume fraction of the material interacts with the different incident e-beam conditions (Supporting Information S11). The Monte Carlo simulations in Figure 5 (a) show that a volume fraction of the fused quartz pillar interacts with the incident electrons when a limited area of the pillar is irradiated with 5-kV, 913-A/m$^2$ (enclosed in the green box) and 30-kV, 1606-A/m$^2$ (enclosed in the purple box) e-beams, respectively. Due to the difference in scattering angle, $\Omega_I$ under a 5-kV, 913-A/m$^2$ e-beam is greater than that under a 30-kV, 1606-A/m$^2$ e-beam. Under a 5-kV, 913-A/m$^2$ e-beam, $f_{\Omega NI}$ (0.19) is 2.53 times less than that under a 30-kV, 1606-A/m$^2$ e-beam (0.48). Consistent with $\Omega_I$ analysis, the fused quartz pillar becomes softer and undergoes greater viscoplastic deformation under a 5-kV, 913-A/m$^2$ e-beam. We confirmed this through *in-situ* compression testing of the fused quartz pillar with a reduced rectangular-shaped area subjected to e-beam irradiation (Figure S13). As shown in the compression load-depth curves in Figure 5 (b), the compression load-depth curves under a 30-kV, 1606-A/m$^2$ e-beam (denoted by purple curves) is higher than that under a 5-kV, 913-A/m$^2$ e-beam (denoted by green curves) at every compression depth. Furthermore, due to $\Omega_I$, there is a significant difference between the compressed shapes of the fused quartz pillars. The SEM images in Figure 5 (c) show that, under a 30-kV, 1606-A/m$^2$ e-beam, intensive viscoplastic deformation occurs around the irradiated area (left SEM image enclosed in purple box). On the other hand, under a 5-kV, 913-A/m$^2$ e-beam, viscoplastic deformation occurs throughout the whole pillar (left SEM image enclosed in the green box). We were able to model these compression load-depth curves and deformed shapes of fused quartz pillars (Figure S14) with the mechanical properties of $\Omega_{NI}$ and $\Omega_I$. We assume that the mechanical properties of $\Omega_I$ under both beam conditions are the same because the current densities are large enough to saturate softening behavior as described in Figure 2(b). Hence, we incorporated these properties into $\Omega_{NI}$ and $\Omega_I$ in the FE model that are estimated based on the Monte-Carlo simulation for each beam condition. Consequently, in Figure 5 (b), the simulated compression load-displacement curves under a 30-kV, 1606-A/m$^2$ e-beam (denoted by purple line with open squares) and a 5-kV, 913-A/m² e-beam (denoted by green line with open squares) match the experimental results quite well. The deformed shape from simulation also describes the plastic deformation concentrated along $\Omega_I$. The deformed shapes under 30-kV, 1606-A/m² and 5-kV, 913-A/m² e-beams (right images enclosed in purple and green boxes in Figure 5 (c)) result from significant viscoplastic deformation concentrated in $\Omega_I$.

For further validation of our simulation approach, we also compared the computational results from the FE model and the experimental results of compression tests on spherical amorphous silica nanoshell. We fabricated through the Stöber process spherical nanoshell structures for which the inside of 280-nm-diameter amorphous silica sphere is hollow and only a 20-nm-thick nanoshell remains. As shown in Figure 5 (d), Monte Carlo simulation suggests that $\Omega_I$ induced under a 1-kV, 6.9-A/m² e-beam covers the entire structure because the shell thickness is small enough to allow the incident electrons accelerated by such low voltage to go through. Hence, when we conducted *in-situ* compression tests (Figure S15), the spherical amorphous silica nanoshell accommodates significant viscoplastic deformation under a 1-kV, 6.9-A/m² e-beam. The spherical nanoshell continuously deforms up to the compression depth of 150 nm (denoted by the red curves in Figure 5 (e)), producing a donut-like shape without any surface cracking (left SEM image enclosed in red box in Figure 5 (f)). In contrast, under beam-off conditions, the deformation of the spherical nanoshell ends with an abrupt load drop (denoted by the black line in Figure 5 (e)) and crack formation on the upper surface of the spherical nanoshell (left SEM image enclosed in the black box in Figure 5 (f)). We also demonstrate that our simulation approach (Figure S16) successfully models this e-beam induced superplastic behavior in deformation of the spherical amorphous silica nanoshell. The compression load-displacement curves from FE simulation with the same mechanical properties of $\Omega_{NI}$ and $\Omega_I$ (black and red lines with open squares in Figure 5 (e)) corresponds well with the experimental results under beam-off and 1-kV, 6.9-A/m² e-beam conditions, respectively (black and red curves in Figure 5 (e)). Furthermore, the deformed shapes from simulation also compare well with the SEM images of the experimentally deformed spheres. Cross-section images show that the spherical nanoshell accommodates deformation through inward bending of the top surface and outward bending of the side surfaces. Under beam-off conditions, the deformation stops with crack formation after minimal bending of these surfaces (right image enclosed in black box in Figure 5 (f)). Under e-beam irradiation, severe bending of theses surfaces is possible, producing the donut-like deformed shape (right image enclosed in red box in Figure 5 (f)).

Lastly, we show that much more complex shaping, similar to the actual glass craft technique in bulk scale, is also possible. We carried out a nanoforging of the spherical amorphous silica nanoshell by pushing it into a nano-trench fabricated on the sapphire substrate with a Pt nano-manipulator. A series of snapshots in Figure 5 (g) clearly shows that the nanoforging of spherical nanoshell is possible under e-beam irradiation. Apparently, without any brittle fracture, the upper and lower surfaces of the spherical nanoshell are deformed along the manipulator and the trench, respectively. In addition, considering $\Omega_I$, we can accurately model this mechanical shaping which is at the level of glass craft. Figure 5 (h) shows how the spherical nanoshell deforms during the nanoforging under e-beam irradiation. The model specifically describes the deformation of the spherical nanoshell by the nano-manipulator and the nano-trench. It suggests that even the complex mechanical shaping of amorphous silica under the e-beam irradiation is predictable and controllable. In this regard, this study can be a breakthrough in the silica glass industries for small-scale devices where the temperature-dependent formability of amorphous silica limits the process compatibility and also the geometry of product. With a proper design of e-beam irradiation, we can improve the formability of silica glass at room temperature and widen its formed geometry compared to the conventional processes such as float glass process, lithography, etching and deposition process.

Therefore, our conceptual mechanical shaping using e-beam can be expanded to a high-throughput manufacturing of silica glass with complex geometries which are becoming essentials in the fields of light-emitting diodes, solar cells, microfluidic devices, and optical fibers. Also, our main finding that the electron-matter interaction can change the nature of strong Si-O covalent bond can be applicable to other covalently bonded non-metallic materials. Accordingly, we believe this study has enormous potential for developing e-beam assisted manufacturing for non-metallic materials.

**Conclusion**

In summary, we investigated the ductile super-plastic deformation of amorphous silica under SEM e-beam irradiation with low $V_A$. Interestingly, the ductility of amorphous silica is not simply proportional to $V_A$ and $J$ of the e-beam. There are optimum $V_A$ and $J$ at which e-beam irradiation maximizes ductility and it relates to the trajectory of the incident electrons and $Ω_I$ generated therefrom inside the amorphous silica. The high energy of the incident electrons induce ionization that changes the interatomic bond nature in the amorphous silica. As a result, even without thermal activation, an amorphous silica sphere can become ductile and deform viscoplastically under e-beam irradiation. Importantly, based on $Ω_I$ analysis, we were able to model the mechanical shaping of various amorphous silica structures under e-beam irradiation. Our models predict the response of the material to external forces, as well as the deformed shapes under e-beam irradiation, well. Accordingly, we have demonstrated the feasibility of mechanically shaping small-scale amorphous silica under e-beam irradiation. We believe that the model can provide an engineering guidance of e-beam induced mechanical shaping that can utilizes e-beam technology widely used in manufacturing processes of brittle silica glass.

**Figure 1**. *In-situ* SEM compression tests on the amorphous silica spheres under a 5-kV, 27.9-A/m$^2$ electron beam. (a) Deformation behavior of amorphous silica spheres in the absence (black) and the presence (light green) of electron-beam irradiation. (a-1) Compression load-depth curves. (a-2, 3, 4, 5, 6) Tilted SEM images of amorphous silica spheres before, during, and after the compression test. (b) Compression load-depth curve under alternating beam-off and beam-on conditions.

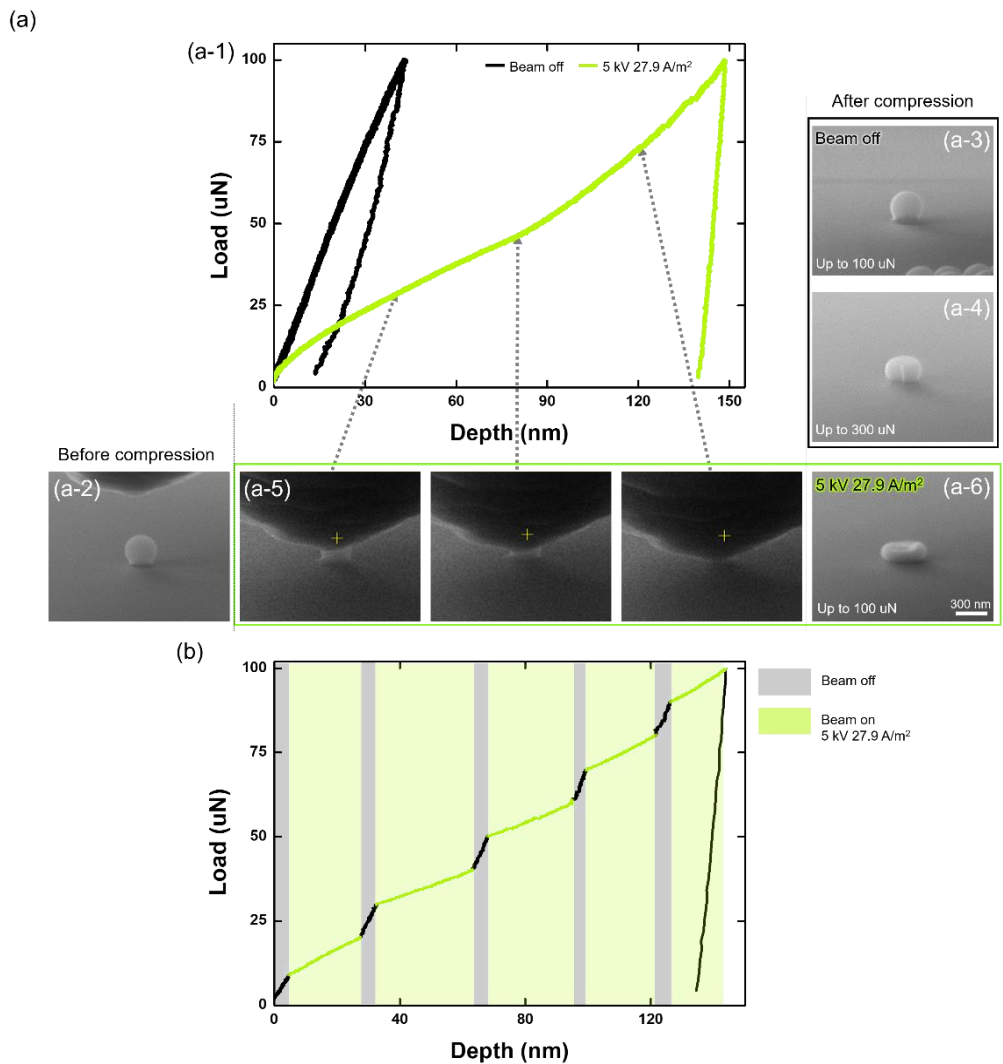

**Figure 2**. Deformation behavior changes of amorphous silica spheres under irradiation using various electron-beam conditions. (a) Tilted SEM images of compressed amorphous silica spheres. The compressed shapes of amorphous silica spheres that differ from the compressed shape obtained under beam-off conditions is enclosed in the red box. (b) Normalized flow stress plot of the amorphous silica spheres compressed under different electron-beam irradiation conditions.

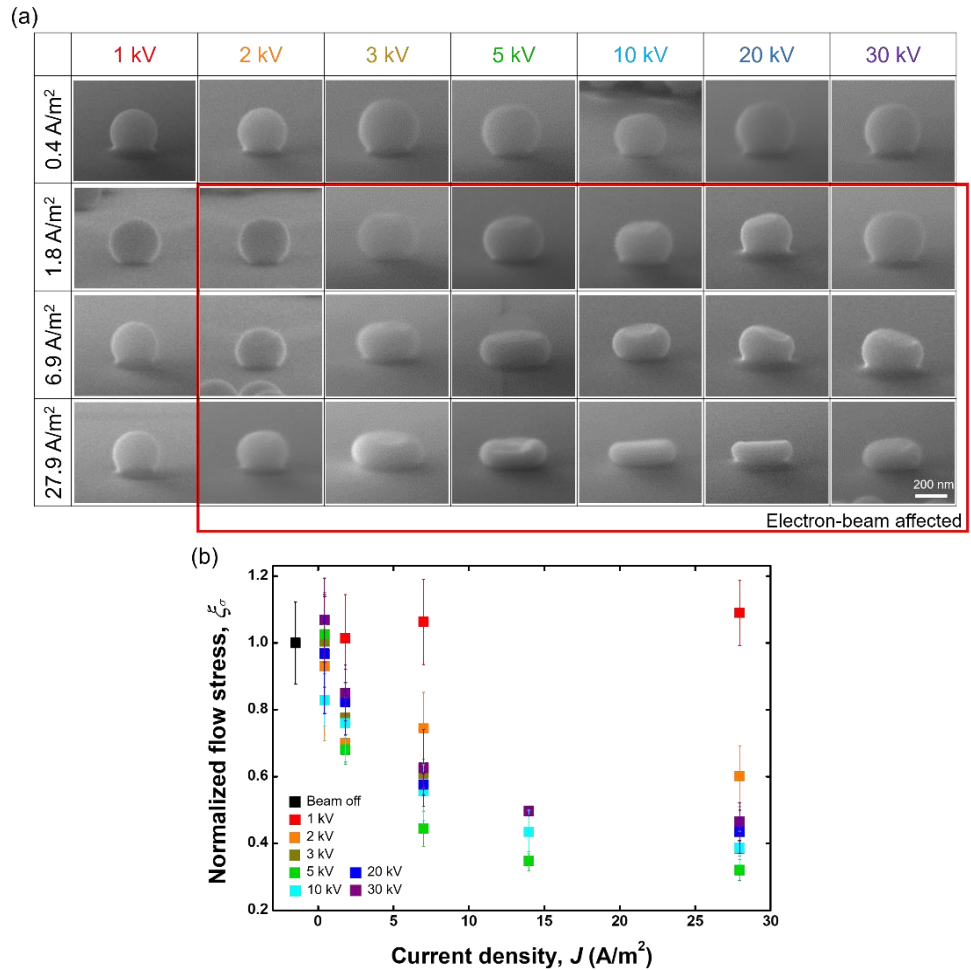

**Figure 3**. Absorbed energy and interaction volume of the irradiated amorphous silica sphere based on a Monte Carlo simulation. (a) Schematic diagram of the simulation model and cross section used for the absorbed energy and interaction volume analysis. (b) Total energy absorbed under different the electron-beam irradiation conditions and (c) absorbed energy distribution in terms of the sphere cross section along (left) and normal (right) to single line scan profile of electron beam. (d) Non-interaction volume fraction under electron-beam irradiation and (e) interaction volume in terms of the sphere cross section along (left) and normal (right) to single line scan profile of electron beam.

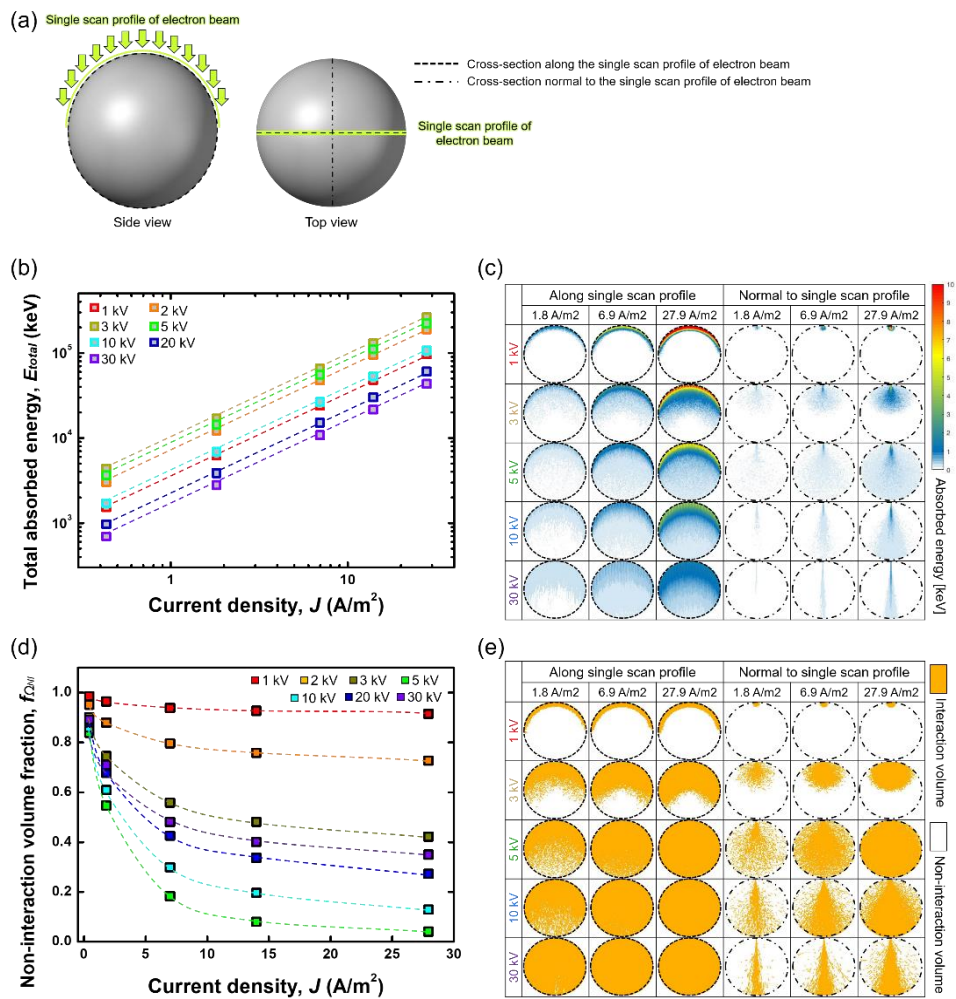

**Figure 4**. Deformation behavior change of fused quartz pillar under various electron-beam conditions. (a) Normalized flow stress of fused quartz pillars deformed under in-situ compression test and corresponding non-interaction volume fractions based on a CASINO Monte Carlo simulation. Inset is a schematic diagram of the simulation model and cross-section used for the interaction volume analysis. (b) Interaction volume in the fused quartz pillar in terms of the cross-section normal to single line scan profile of electron beam. (c) Analysis of strain rate sensitivity and (d) Activation volume of the fused quartz pillar during compression under various electron-beam irradiation conditions.

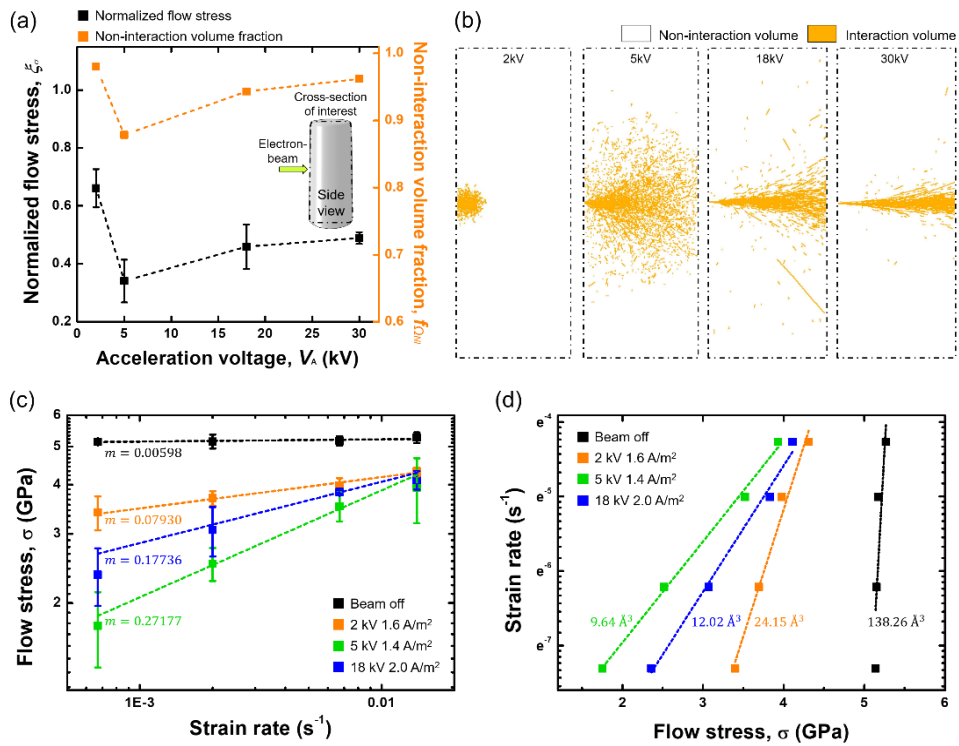

**Figure 5**. Mechanical shaping of (a-c) fused quartz pillars and (d-f) spherical amorphous silica nanoshells by controlling interaction volume. Electron-beam conditions during mechanical shaping are: beam-off (black) 1 kV, 6.9 A/m$^2$ (red), 5 kV, 913 A/m$^2$ (green), and 30 kV, 1606 A/m$^2$ (purple). (a, d) Interaction volume within amorphous silica structures under each set of electron-beam conditions. Schematic diagram of interaction model (top) and cross-section view of interaction volume (bottom). (b, e) Compression load-depth curves of amorphous silica structures obtained experimentally (lines) and through simulation (lines with open squares). (c, f) Compressed shape of amorphous silica structures obtained experimentally (left) and through simulation (right). Nanoforging of spherical amorphous silica nanoshell from in-situ experiment (g) and FE model (h).

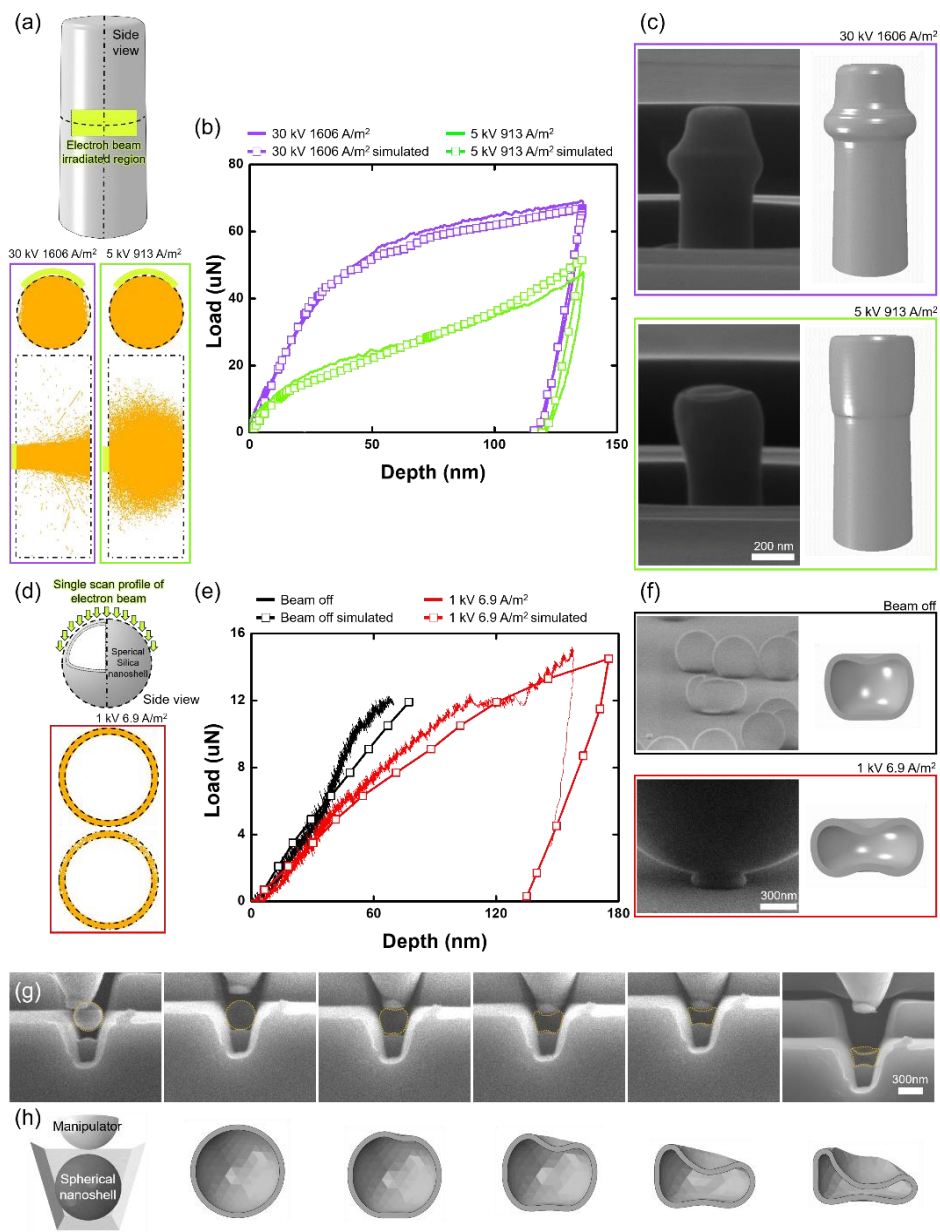